\documentclass{IEEEcsmag}
\usepackage[colorlinks,urlcolor=blue,linkcolor=blue,citecolor=blue]{hyperref}
\usepackage{upmath}
\usepackage{graphicx}
\usepackage{epsfig}
\usepackage{epstopdf}
\usepackage{float}
\jvol{XX}
\jnum{XX}
\paper{8}
\jmonth{}
\jname{IEEE Multimedia}
\pubyear{2022}

\begin{document}

\title{Interpretability of Machine Learning: Recent Advances and Future Prospects}

\author{Gao, Lei}
\affil{Toronto Metropolitan University}

\author{Guan, Ling}
\affil{Toronto Metropolitan University}

\author{}
\affil{}
\begin{abstract}
\small
  The proliferation of machine learning (ML) has drawn unprecedented interest in the study of various multimedia contents such as text, image, audio and video, among others. Consequently, understanding and learning ML-based representations have taken center stage in knowledge discovery in intelligent multimedia research and applications. Nevertheless, the black-box nature of contemporary ML, especially in deep neural networks (DNNs), has posed a primary challenge for ML-based representation learning. To address this black-box problem, the studies on interpretability of ML have attracted tremendous interests in recent years. This paper presents a survey on recent advances and future prospects on interpretability of ML, with several application examples pertinent to multimedia computing, including text-image cross-modal representation learning, face recognition, and the recognition of objects. It is evidently shown that the study of interpretability of ML promises an important research direction, one which is worth further investment in.
\end{abstract}
\maketitle
\section{I. INTRODUCTION}
In recent years, machine learning (ML), especially deep neural networks (DNNs) and artificial intelligence (AI) in general, have been utilized broadly and successfully in different multimedia computing tasks, such as audio processing, image classification, computer vision, image retrieval, and healthcare, amongst others \cite{1}-\cite{2}. It is widely acknowledged that such tasks involve the processing of various information streams to gain valuable insights from the input data sources, intermediate decisions, or higher level activities, leading to superior, sometimes unprecedented performance. Despite the extraordinary success of ML and AI in multimedia and other fields which require intelligent processing, the interpretability of ML/AI remains a persistent challenge. Specifically, the black-box nature of contemporary ML architectures has posed a longstanding problem, causing concerns about questionable performances and predictions in real applications \cite{3}. Therefore, there has been urgent demand to better understand and more effectively learn ML-based representations. To address this black-box problem, interpretable ML (I-ML) methods have recently drawn considerable attention and interests in ML and the intelligent multimedia communities \cite{4}-\cite{5}.
\begin{figure}[H]
\centering
\includegraphics[height=1.0in,width=3.2in]{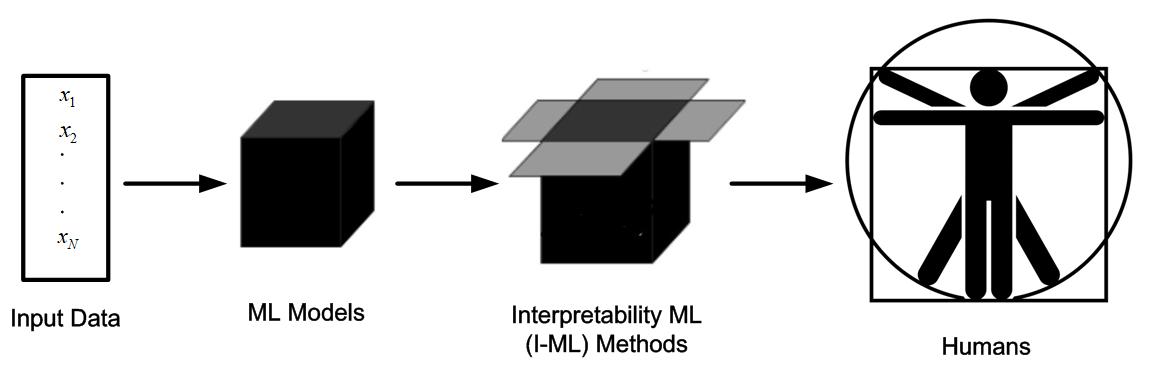}\\ Figure 1 The interpretable ML (I-ML).
\end{figure}
An illustrative diagram of I-ML is presented in Figure 1 \cite{4}. %with the spectra of interpretability/explainability shown in Figure 2 \cite{5}. In Figure 2, while the explainable models on the right attempt to explain what happens in a black-box in ML, those on the left are, at least partially, interpretable using analytical, mathematical and/or experimental methods.
As consensus suggests \cite{80}-\cite{81}, the classical neural network (NN)-based models (e.g., neural network, convolutional neural network (CNN) and DNNs in general) exhibit less interpretable characteristics, thus attracting more attention from both academic and industrial sectors, first attempting to explain the black-box and, more recently, designing new models that are inherently interpretable.\\\indent
%\begin{figure*}[t]
%\centering
%\includegraphics[height=0.8in,width=4.2in]{figure1.jpg}\\ Figure. 2 The interpretability spectrum in ML \cite{5}.\\
%\end{figure*}
Though all NN-based models stem from Kurt-Vladimir (K-V) Universal approximation (UA) theory \cite{6}-\cite{7}, deep learning (DL) has dominated the research landscape for the past 10 years in visual computing, natural language processing, video processing, and more \cite{8}-\cite{11}. There are several reasons for the booming popularity of DL-based models: the drastically increased chip processing abilities (e.g. GPU units), the significantly lowered cost of computing hardware, the considerable advances in ML \cite{12} and the knowledge in neurobiological science discovered and accumulated over several decades \cite{72}. In general, DL-based architectures consist of multiple processing layers to learn representations of input data with different levels of abstraction. Combined with certain optimization algorithms (e.g., backpropagation, Adam, etc.), such architectures help reveal the intricate structure in large data sets to develop intelligent systems/machines which attempt to mimic the natural human computing system for information processing. Nevertheless, the end-to-end architecture usually makes the DL-based representations a black-box \cite{13}\cite{14}, implying that it is difficult to tell what the prediction relies on, and what features or representations play more important roles in a given task. For a similar reason, DNNs have been shown to suffer from lack of robustness \cite{15}. For example, small changes in an input, sometimes imperceptible to humans, could induce instability in the DNNs, causing undesirable performance \cite{15}. In the last few years, the ML community recognized that either the black-box problem has to be understood and solved or truly interpretable models have to be conceived for better design and development of the ML models. Consequently, the study of explainable and interpretable ML models came into play. Though the objectives of explainability and interpretability are similar, they take very different approaches. Explainability of ML refers to the capability of understanding the work logic of a ML model after the completion of its design \cite{5}, thus attempting to solve the black-box problem by post hoc actions \cite{95}. On the other hand, interpretability of ML usually refers to the ability that users can not only see but also study and understand how inputs are mathematically and/or logically mapped to the outputs \cite{5}. The ultimate goal of studying interpretability is to build the model architecture, which is inherently interpretable to avoid the black-box problem \cite{94}. Either way, the objective of I-ML is to identify the best possible strategies to improve interpretability in intelligent multimedia, and ML in general. For reference, the number of articles on interpretability/explainability in the period of 2000-2019 is plotted in Figure 2  \cite{16}.
\begin{figure}[H]
\centering
\includegraphics[height=1.61in,width=2.2in]{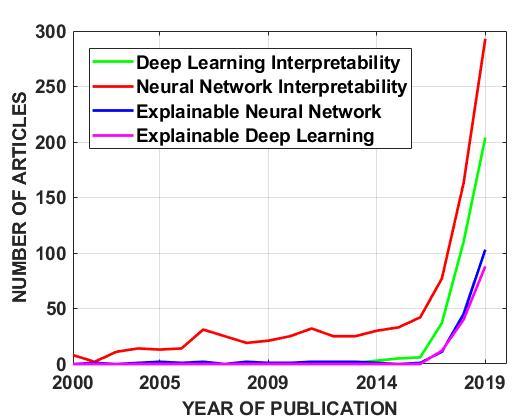}\\ Figure 2 The number of articles on interpretability and explainability of NNs/DL \cite{16}.
\end{figure}
As commonly agreed upon, there are two steps in making NN/DL models interpretable: extraction and exhibition \cite{15}. In general, extraction discovers relevant knowledge for an intermediate representation, and exhibition organizes such representation in a way that is easy for humans to understand, e.g., via visualization of the representation \cite{15}. The core of this survey focuses on finding relationships either contained in the data or learned by the ML model, thus more inline with the extraction step of I-ML. Several survey papers have been made available \cite{15}-\cite{19}\cite{35}\cite{36}\cite{73}-\cite{75} to the ML and intelligent multimedia communities, mostly focusing on explaining the internal structure of a black-box. While this article will further dig into recent advances along this line of research, emphasis will be given to another class of models which are designed to be inherently interpretable from analytically inspired perspectives.\\\indent To complement the review, a down-to-earth approach is utilized in this work, applying the I-ML methods to three multimedia related areas (text-image representation, face recognition and object recognition), in anticipation that such an approach would better inspire readers in pursuit of ML interpretability from multiple angles in their R \& D endeavors. In the rest of this paper, Sections II and III present reviews on the classical NN-based and inherently interpretable models, respectively. In Section IV, the evaluation and comparisons of representative models pertinent to multi-modal image and multimedia analysis and recognition are conducted. Section V presents a summary discussion and outlines some future prospects. Section VI draws conclusions.
%As commonly agreed upon, there are two steps in making NN/DL models interpretable: \emph{extraction} and \emph{exhibition} \cite{15}. Specifically, the \emph{extraction} contains an intermediate representation, and the \emph{exhibition} generates intermediate representations in a way that is easy for humans to understand.
\section{II. Classical NN Based Methods}%Interpretability of NN By FFNN or Pure DL Based Methods}
In the past several years, classical NN-based models have achieved great success and outperform humans in numerous difficult tasks, such as image and visual classification, natural language processing and recognition, and board games. However, the black-box nature of the classical models presents a real challenge to explain the underlying mechanisms and behaviors of the networks. The popular solution to address this problem is to explore interpretability via explaining the inside of the black-box. Hence, for this class of models, the word `interpretability' refers to the ability to clarify and extract knowledge representations in different layers of NN-based models as defined in \cite{20}. In this section, the most studied methods in this class, FFNN based and DL based, are surveyed.
\subsection{FFNN Based Methods}
Back in the 1980's, the FFNN was already employed to interpret and design NNs and other highly sophisticated networks \cite{15}. For instance, a FFNN was utilized to construct a global minimum loss function, resulting in a certain level of understanding of the model \cite{21}. In \cite{22}, Kuo \emph{et al}. presented an interpretable feedforward (FF) design using a data-centric approach, by which the network parameters of the current layer are derived according to data statistics from the output of the previous layer in a one-pass manner. Yosinski \emph{et al.} \cite{34} inspected the activation values of neurons in each layer with respect to different images or videos. They found that live activation values that change for different inputs are helpful for understanding how a given model works, thereby generating an interpretable model.
\subsection{DL Based Methods}
Due to the recent advancement of DNNs, a plethora of works based on pure DL methods have been proposed, forming the mainstream tactics in identifying explainability of DNNs, especially for CNNs. For example, Zhang \emph{et al}. \cite{20} proposed interpretable CNNs to clarify knowledge representations in high convolution layers. Then, the generated knowledge representation aids in the understanding of logic inside a CNN architecture. Zee \emph{et al}. \cite{23} introduced an interpretable neural network (e.g., Siamese CNN) with application to face recognition. This network provides an explainable model to better distinguish between the faces of two similar actresses. In \cite{26}, a technique called CNN-INTE is introduced and applied to explain deep CNNs. The CNN-INTE method provides global interpretation for any test instances on the hidden layers in the whole feature space, attempting to explain the inner mechanisms of DL-based models. Chen \emph{et al.} \cite{29} proposed a prototype layer that was then added to a regular CNN architecture. According to the prototype layer, the network can provide several prototypes for different parts of the input image, resulting in proper interpretation of the model's functionality. In \cite{30}, a decision tree is presented to encode decision modes in fully-connected layers. Instead of classification, the decision tree is designed to quantitatively explain the logic for each CNN prediction. Wang \emph{et al.} \cite{31} associated every output channel in or from each layer with a gate (non-negative weight) for CNNs, which indicates how critical that channel is in the overall network architecture. By doing so, the network gains the ability to explore and assign meanings to critical nodes so that the CNNs become explainable. Ribeiro \emph{et al.} \cite{33} introduces Local Interpretable Model-Agnostic Explanations (LIME), an algorithm that provides explanations of decisions for any ML model. The LIME algorithm calculates the importance of each feature by generating perturbed samples of the input point and using these samples (labeled by the original model) to learn a local approximation to the CNN model.\\\indent In addition, a grouping-based interpretable neural network (GroupINN) \cite{24} was proposed. By utilizing three different types of layers: node grouping layer, graph convolutional layer and fully connected layer, GroupINN can learn the node grouping and extract graph features jointly, leading to improved performance on brain data classification. Montavon \emph{et al.} \cite{25} proposed a layer-wise relevance propagation (LRP) technique, specifically designed for explanation of CNNs. In essence, the LRP method is rooted in a conservation principle, where each neuron receives a share of the network output and redistributes it to its predecessors in equal amount until the input variables are reached, an operation procedure similar to the autoencoder algorithm. Hooker \emph{et al.} \cite{27} presented a RemOve And Retrain (ROAR) method to evaluate DL-based interpretability, which is accomplished by verifying how the accuracy of a retrained model degrades as features estimated to be important are removed. In \cite{28}, the Locality Guided Neural Network (LGNN) method is proposed with application to explainable artificial intelligence (XAI). Since LGNN is able to preserve locality between neighbouring neurons within each layer of a deep network, it can alleviate the black-box nature of current AI methods and make them understandable by humans to some extent. In \cite{32}, an interpretable partial substitution (a rule set) is investigated on the network to cover a certain subset of the input space. The proposed model integrates an interpretable partial substitute with any black-box model to introduce transparency into the predictive process at little to no cost. \\\indent The schematic graph of the interpretability by FFNN/DL is depicted in Figure 3. In essence, the power of the DL-based methods reviewed in this section is confined by certain limitations when applied to explore NN interpretability such as the vanishing/exploding gradient problem, and tuning of parameters manually. These limitations prompt several groups of researchers to consider investigating model interpretability from alternative angles, which is the focus of the next section.
\begin{figure}[H]
\centering
\includegraphics[height=1.2in,width=2.0in]{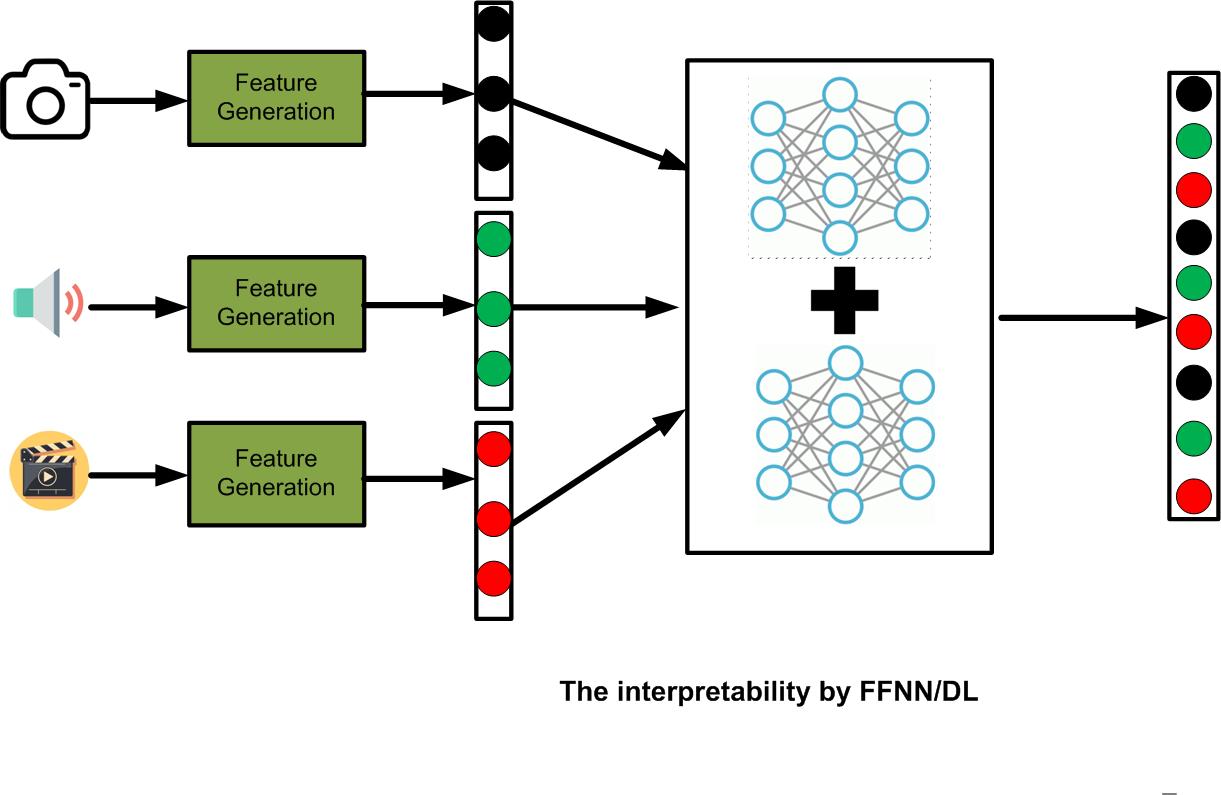}\\ Figure 3 The interpretability by FFNN/DL.
\end{figure}
\section{III. Inherently I-ML Methods}%Interpretability of NN By Integration of SGO with NN Architecture}
The second class of I-ML models, coined as inherently interpretable models, obeys structural knowledge of the domain, such as monotonicity, causality, structural (generative) constraints, additivity, or physical constraints that come from domain knowledge and can, at least, be partially justified by theoretical analysis such as physics laws and/or mathematical formulas \cite{73}. The key members of this I-ML family include physics-informed, model based, algorithm unrolling solutions and mathematics inspired methods, which will be presented in the following subsections.
\subsection{Physics-informed NN}
In general, methods in physics-informed NN are trained to handle supervised learning tasks while respecting any given laws of physics described by general nonlinear partial differential equations \cite{69}. The schematic of a physics-informed NN is given in Figure 4, where a fully-connected neural network is adopted to approximate the multi-physics solutions $u$. The derivatives of $u$ are computed with automatic differentiation (AD) and then utilized to formulate the residuals of the governing equations in the loss function. Finally, parameters of the neural network $\theta$ and the unknown partial differential equation (PDE) parameters $\lambda$ are studied by minimizing the loss function \cite{82}--\cite{83}.
\begin{figure}[H]
\centering
\includegraphics[height=1.2in,width=2.8in]{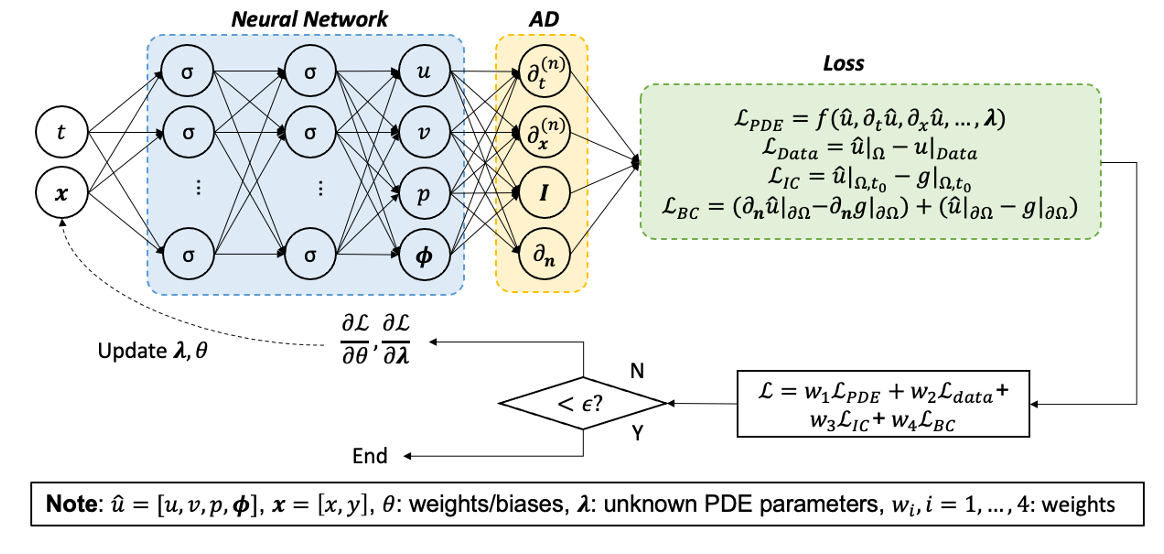}\\ Figure 4 A schematic of a physics-informed NN model from \cite{82}.
\end{figure}
There exist several representative physics-informed models. Raissi \emph{et al.} \cite{84} utilized the physics-informed NN to solve two problems in ML: data-driven solution and data-driven discovery of partial differential equations, reporting promising results for a diverse collection of problems in computational science. In \cite{85}, a fractional physics-informed NN model was proposed to address multidimensional forward and inverse problems with forcing terms whose values are only known at randomly scattered spatio-temporal coordinates (black-box forcing terms). Cuomo \emph{et al.} \cite{86} published a survey paper which summarized that most research in physics-informed NN has focused on customizing this class of models through different activation functions, gradient optimization techniques, neural network structures, and loss function structures in ML.
\subsection{Model based NN}
Studies on interpretability of model-based NN mainly focus on the construction of models that readily provide insight into the relationships they have learned \cite{70}. A model based NN diagram is shown in Figure 5. In the diagram, there exist three streams according to different domain knowledge, the purely data-driven model (the left stream), model-based ML (the middle stream), and model-based algorithm without data-driven characteristics (the right stream) \cite{87}.
 \begin{figure}[H]
\centering
\includegraphics[height=1.4in,width=3.0in]{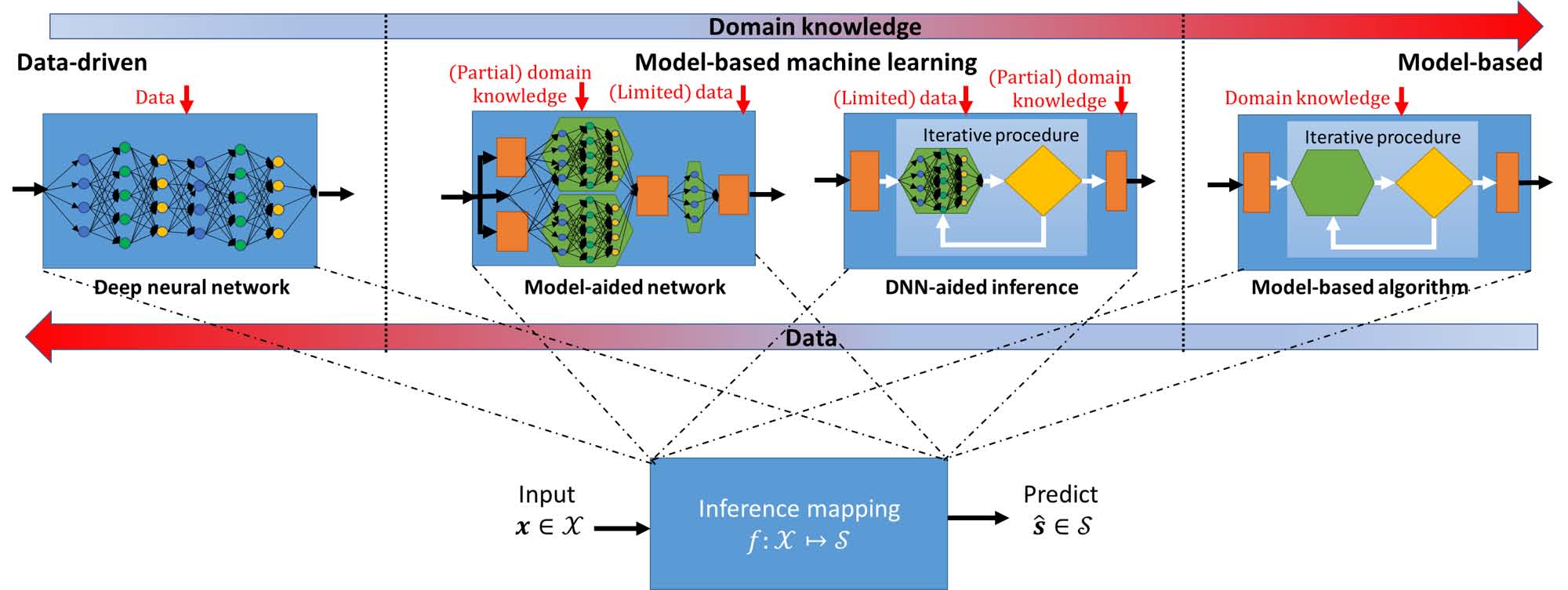}\\ Figure 5 A diagram of model based NN from \cite{87}.
\end{figure}
Model based architecture has been adopted for DNN research and applications. In \cite{88}, a model based NN framework was presented for image reconstruction. This framework provides a systematic approach for deriving deep architectures for inverse problems with an arbitrary structure, generating an interpretable DNN model for a variety of image applications. Shlezinger \emph{et al.} \cite{89} utilized the integration of model based NN with data-driven pipelines to introduce a general framework for deep learning. The framework can be applied to a broad range of application areas, including ultrasound imaging, optics, digital communications, and tracking of dynamic systems. In \cite{90}, a model based NN was proposed for optimized sampling and reconstruction. This algorithm facilitates the joint and continuous optimization of the sampling pattern and the CNN parameters to improve image quality, exhibiting certain levels of interpretability.
\subsection{Algorithm Unrolling}
Different from \cite{69}\cite{70}, algorithm unrolling solves model interpretability by providing a concrete and systematic connection between iterative algorithms that are widely used in signal processing and DNNs \cite{71}. A high-level overview of algorithm unrolling is drawn in Figure 6. From Figure 6, given an iterative algorithm (left), a corresponding deep network (right) is generated by cascading its iterations $h$. Then, iteration step $h$ (left) is executed a number of times, resulting in different parameters $h^1$, $h^2$,...(right). Each iteration $h$ depends on algorithm parameters $\theta$, which are transferred into network parameters ${\theta}^1$, ${\theta}^2$,...(right). Instead of determining parameters through cross-validation or analytical derivations, the parameters ${\theta}^1$, ${\theta}^2$,... are learned from training datasets through end-to-end training. In this way, the network layers naturally inherit interpretability from the iteration procedure.
\begin{figure}[H]
\centering
\includegraphics[height=1.1in,width=2.8in]{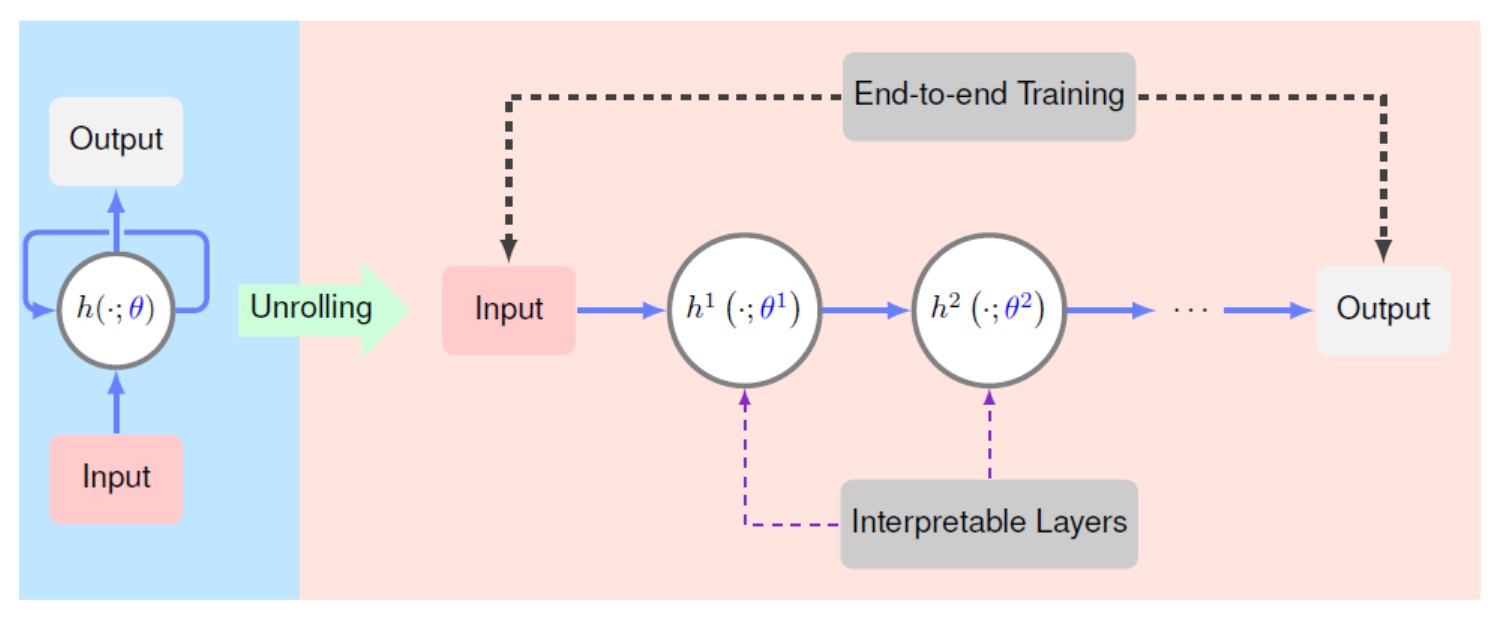}\\ Figure 6 A high-level overview of algorithm unrolling from \cite{71}.
\end{figure}
In \cite{91}, a Bayesian based unrolling algorithm was presented with application to single-photon Lidar systems. The resulting algorithm benefits from the advantages of both statistical and learning based frameworks, providing improved network interpretability. Chen \emph{et al.} \cite{92} proposed a graph unrolling network algorithm for graph signal denoising. The proposed method expanded the original unrolling algorithm  to the graph domain and provided an interpretation of the architecture design from a signal processing perspective. In \cite{93}, an interpretable unsupervised unrolling algorithm was introduced to hyperspectral pansharpening. In this work, a pansharpening model was first constructed. Then, iterative steps are unfolded into a deep interpretable iterative generative dual adversarial network.
\subsection{Mathematics Inspired Methods}
%Amongst this branch of inherently I-ML models, canonical correlation analysis (CCA), stands out, exhibiting tremendous potential to address the interpretability challenge in ML research. By integrating statistical machine learning (SML) principles, coined as
By integrating Statistics Guided Optimization (SGO) with NN architecture, this class of models, coined as SGO-NN, exhibits model agnostic properties and is ideal for global model interpretability. In fact, the idea also stems from the K-V UA theory \cite{6}-\cite{7}, but using an alternate realization strategy. Instead of going deep, the network goes wider, an approach first acknowledged and practiced by the NN community in the 1980's. Like the DNNs, computational power limited the progress of this class of networks as well until the first decade of the 21st century. However, DNNs quickly dominated the ML landscape, and thus overran further exploration of this alternative approach until recently.\\\indent This class of architecture features three characters: a) Kolmogorov-Arnold (K-A) theorem \cite{79} and K-V UA theory \cite{6}-\cite{7} as the foundation; b) certain biological justifications and scientific rules in architecture design; c) powerful optimization methods for a quality training process. Analytically, the recent progress in approximation theory solidly verified the K-A theorem/K-V UA theory that three hidden layers are sufficient for a NN to approximate any nonlinear functions under mild conditions \cite{38}. At the same time, a number of practical models with three or fewer layers have emerged and demonstrated their flexibility \cite{39}, effectiveness and computational affordability. Examples include PCANet \cite{40}, DCTNet \cite{41}, CCANet \cite{42}, DDCCANet \cite{43}\cite{67}, ILMMHA \cite{44}. Ma \emph{et al}. \cite{40} offered a PCANet for image classification. In the PCANet, a classical SGO method, principal component analysis (PCA), was utilized to construct multistage filter banks, leading to easy understanding of the proposed model. In \cite{41}, DCTNet was proposed by integrating discrete cosine transform (DCT) with a NN architecture, resulting in an analytically interpretable model. In \cite{42}, a canonical correlation analysis network (CCANet) is presented. The CCANet constructs two-view multistage filter banks by employing the canonical correlation analysis (CCA) algorithm and designs a NN architecture with interpretable properties. In order to explore more discriminant information from given data sets, the within-class and between-class correlation matrices are employed and optimized jointly to construct the convolution layer, laying the foundation for the introduction of a distinct discriminant canonical correlation analysis network (DDCCANet) \cite{43}\cite{67}. In \cite{44}, a learning-based multi-modal hashing analysis (ILMMHA) model is proposed. ILMMHA is able to generate an analytically interpretable feature representation, yielding substantially improved performance in cross-model (text-image) recognition application.\\\indent It is worth noting that the key members of this model class, such as CCANet, DDCCANet and ILMMHA, are particularly prevalent to information processing by mimicking certain facts in neurobiological signal analysis, handling multiple information streams coherently and simultaneously (with audio-visual processing as a popular example) \cite{45}. Apparently such an architecture fits well with multimedia information processing, in which two or more different data streams are processed jointly. It is worth further noting that even when working with one mode of sensory data, e.g., audio or visual, such a principle still applies. For instance, a) human speech is a natural blend of phonetic information and vocal information; b) to form a 3D color image in the human visual system, color and depth information are jointly processed and presented.\\\indent While K-A theorem/K-V UA theory serve as the theoretical foundation and facts of neurobiology helps form the architecture, learning other than steepest descent motivated backpropagation (Back-Prop) algorithms has also been approached. Though it is a known fact that steepest descent is slow to converge (and easy to diverge) and prone to get trapped in local minima, it forms the backbone of Back-Prop algorithms due to the very limited computational power available in the 1980's when the algorithm emerged. The extremely long training time for contemporary DL algorithms, e.g., in any Resnet architecture, is, at least, partly due to the above mentioned issue. Many ideas have been put forward to alleviate this problem. For example, Widrow \emph{et al.} \cite{37} proposed a Non-Prop algorithm and demonstrated that it is much simpler and easier to implement, and converges much faster than Back-Prop algorithms while producing similar quality of results for shallow NNs. Different from Widrow's idea, the multi-stream nature of neurobiological signal processing has recently motivated a different Non-Prop algorithm in which the parameters of the network are determined by analytically solving a SGO problem in each convolutional layer independently, leading to model interpretability which is mathematically plausible. Therefore, the mathematical vigor and meaningful interpretation of the functionality of the SGO solutions offer an effective vehicle to complement NNs' ability to handle big data and have inspired researchers in ML and multimedia communities to make collective use of their strengths to explore the model interpretability. Note that, since this class of NN models executes the optimization process by utilizing SGO-based algorithms instead of Back-Prop, it shortens the calculation process and reduces running time, making SGO-NN algorithms more applicable in real tasks. The schematic graph of such a network architecture is illustrated in Figure 7. %Furthermore, since the visualization of convolution filters is considered as another direct way of exploring NN interpretability \cite{15}, a great number of methods using SML principles with NN architecture are proposed as another class of analytically interpretable models which is able to visually explain the functionalities of convolution filters (e.g., PCANet \cite{16}, DCTNet \cite{17}, etc.), resulting in interpretable models. %Specifically, Ma \emph{et al}. \cite{16} offered a PCANet for image classification. In the PCANet, a classical SML method, principal component analysis (PCA) was utilized to construct multistage filter banks, leading to interpretability of the proposed model. In \cite{17}, DCTNet was proposed by integrating discrete cosine transform (DCT) with a NN architecture. In DCTNet, since the two dimensional (2D) DCT is employed to generate the filters for each convolution layer followed, the DCT-based knowledge representations are generated in different layers, resulting in an analytically interpretable model.
\begin{figure}[H]
\centering
\includegraphics[height=1.1in,width=2.2in]{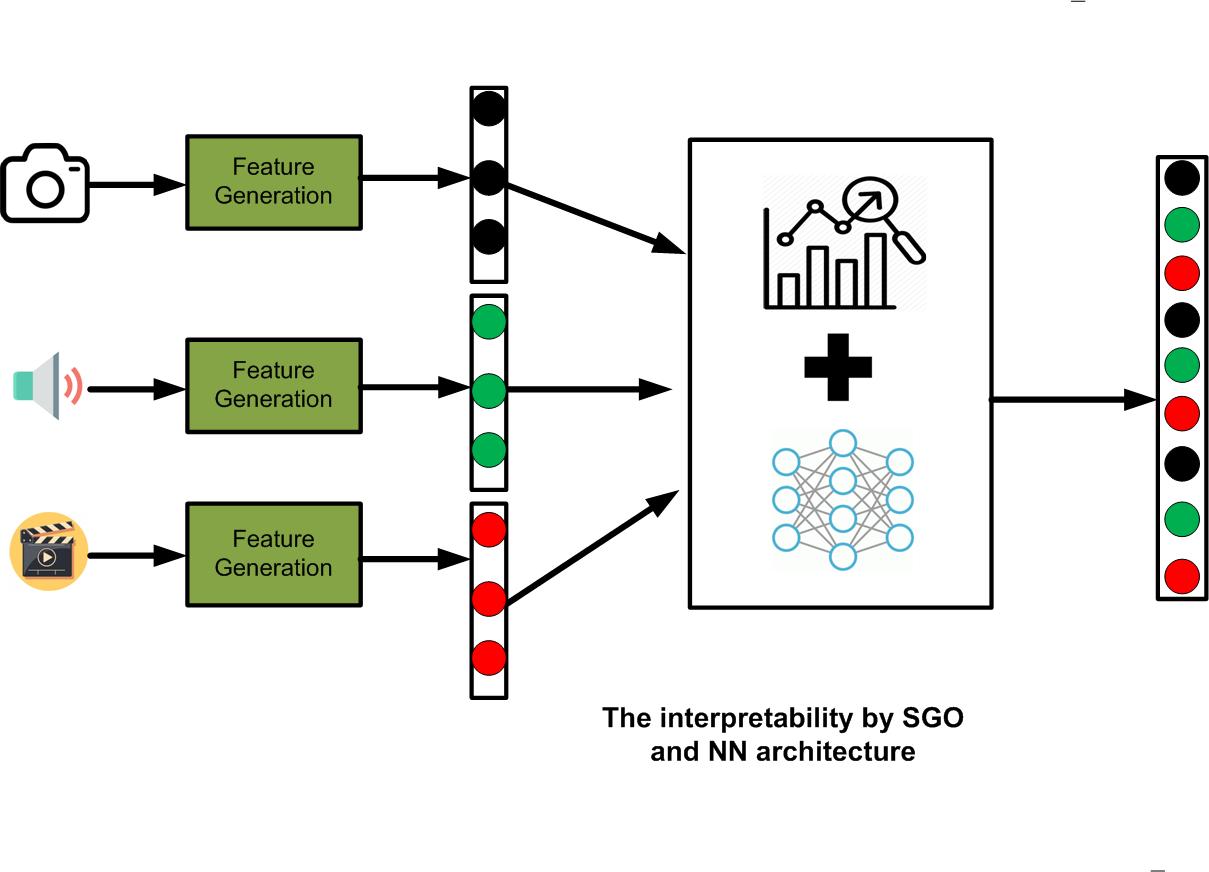}\\ Figure 7 The interpretability by SGO and NN architecture.
\end{figure}
\section{IV. Evaluation on Exemplar Applications}
ML techniques have been playing a significant role in information processing with impressive, sometimes unprecedented success. In the following subsections, the performances of different state-of-the-art (SOTA) I-ML methods pertinent to multi-modal image and multimedia analysis and recognition are evaluated on cross-modal (text-image)-based and multi-view visual-based (face recognition and recognition of objects) examples. The compared algorithms/models are grouped into three categories, a) without I-ML (WO-I-ML), b) with classical NN (C-NN), and c) SGO-NN. Note that, all SGO-NN models used in the experiments consist of three or fewer hidden layers.
\subsection{\textbf{Cross-modal (Text-image) Recognition}}
\subsection{\emph{The Wiki Database.}}
The Wiki database is captured from various featured articles in Wikipedia. There are 2,866 documents stored in text-image pairs and associated with supervised semantic labels of 10 classes. For fair comparison, all documents are further divided into a training subset with 2173 documents
and a testing subset with 693 documents as practiced in past studies \cite{46}-\cite{50}. The SGO-NN model, ILMMHA, uses two classical features as the input (bag-of-visual SIFT (BOV-SIFT) \cite{48} and the Latent Dirichlet Allocation (LDA) \cite{49}).
%Then, features are extracted using the bag-of-visual SIFT vector (BOV-SIFT) from images \cite{48} and the Latent Dirichlet Allocation (LDA) vector from texts \cite{49} with performance given in \textbf{Table 1}. Next, ILMMHA is applied to the BOV-SIFT and LDA features with the result shown in \textbf{Table 1}.
For comparison, the recognition results by different types of SOTA algorithms are tabulated in the \textbf{Table 1}.\\
\vspace*{-10pt}
\begin{table}[h]
\small
\renewcommand{\arraystretch}{1.0}
\caption{\normalsize{Recognition accuracies on the Wiki database}}
\setlength{\abovecaptionskip}{0pt}
\setlength{\belowcaptionskip}{10pt}
\centering
\tabcolsep 0.07in
\begin{tabular}{cccc}
\hline
\hline
Methods & Training \# & Accuracy & Type\\
\hline
\hline
%BOV-SIFT \cite{48} & 2173 & 17.21\% &WO-I-ML\\
%LDA \cite{49} & 2173 &63.44\% &WO-I-ML\\
%PLS \cite{92} & 2173 & 48.14\%\\
%DCML \cite{93} & 2173 & 55.40\%\\
%CCA \cite{50} & 2173 &63.64\%&WO-I-ML\\
%LMCCA \cite{46} & 2173 &64.79\%&WO-I-ML\\
%DMCCA \cite{47} & 2173 &65.22\%&WO-I-ML\\
%SCM-Seq \cite{94} & 2173 &63.35\%&W-I-ML\\
%MSC \cite{95} & 2173  &61.48\% \\
%MVLS \cite{96} & 2173 &62.70\%\\
$L_{2,1}$CCA \cite{46} & 2173 &65.99\%&WO-I-ML\\
%SPIB \cite{52} & 2173 &65.50\%&WO-I-ML\\
MH-DCCM \cite{47} & 2173 &67.10\%&WO-I-ML\\
RE-DNN \cite{50} & 2173 &63.95\% &C-NN\\
\bf{ILMMHA} \cite{44} & 2173 & \bf{74.28\%}&SGO-NN\\
\hline
\hline
\end{tabular}
\end{table}\\\indent
Since the incumbent SGO principle (multi-modal hashing (MMH)) is capable of measuring semantic similarity across multiple variables effectively, the ILMMHA model yields a new feature representation of high quality by integrating the SGO principle with NN architecture. \textbf{Table 1} shows that ILMMHA outperforms the others by a large margin, demonstrating the superiority of this SGO-NN model.
\subsection{\textbf{Visual Examples}}
\subsection{\emph{Face Recognition-The ORL Database.}}
In face recognition, experiments are conduced on the Olivetti Research Lab (ORL) database. In the ORL database, there are 40 people with 10 different images for each subject. The images are captured under different conditions, such as illumination, posing, etc. In this experiment, all 400 samples in the ORL database are utilized with 280 images randomly selected as the training samples while the remaining images used in testing. The SGO-NN models are worked on two-view data sets (the original image and local binary patterns (LBP)-based image).
%For CCANet and DDCCANet, a visual feature (local binary patterns (LBP)) is performed on each sample to generate the second view data set.
%The schematic diagram of SGO-NN models on the ORL database is shown in Figure. 5.
The comparison amongst SOTA algorithms is given in \textbf{Table 2}.
%\begin{figure}[H]
%\centering
%\includegraphics[height=1.4in,width=3.0in]{Fig3.jpg}\\ Figure. 5 The diagram of SGO-NN models on the ORL database.
%\end{figure}
\vspace*{-10pt}
\begin{table}[h]
\small
\renewcommand{\arraystretch}{1.0}
\caption{\normalsize{Recognition accuracy on the ORL database}}
\setlength{\abovecaptionskip}{0pt}
\setlength{\belowcaptionskip}{10pt}
\centering
\tabcolsep 0.05in
\begin{tabular}{cccc}
\hline
\hline
Methods & Training \#  & Accuracy & Type\\
\hline
\hline
%MCPCADP \cite{20} & 200 & 91.25\%\\
%MDL \cite{21} & 200 &92.15\%\\
%WLCRC \cite{22} & 200 & 92.69\%\\
%GLS \cite{23} & 200 & 93.33\%\\
%GDLMPP \cite{55} & 200 &94.50\%&WO-I-ML\\
%CCA \cite{50} & 200 &94.50\% &WO-I-ML\\
%MCCA \cite{56} & 200 &94.50\% &WO-I-ML\\
%DSR \cite{57} & 200  &95.00\% &WO-I-ML\\
%SOLDE-TR \cite{58} & 200 &95.03\%&WO-I-ML\\
%CDPL \cite{59} & 200 &95.42\%&WO-I-ML\\
%AOS+Custom VGG \cite{60} & 200 &93.62\%&FFNN-DL\\
%CCANet \cite{42} & 200 &94.70\%&SGO-NN\\
%\bf{DDCCANet} \cite{43} & 200 & \bf{95.80\%}&SGO-NN\\
%\hline
%\hline
ANFIS-ABC \cite{51} & 280 &96.00\%&WO-I-ML\\
ESP \cite{52} & 280 & 96.00\%&WO-I-ML\\
DL-SE \cite{76} & 280 & 96.08\%&WO-I-ML\\
HMMFA \cite{77} & 280 & 94.17\%&WO-I-ML\\
CNN \cite{53} & 280 &95.92\%&C-NN\\
IKLDA+PNN \cite{54}  & 280 &96.35\%&C-NN\\
LiSSA \cite{55} & 280 & 97.51\%&C-NN\\
PCANet \cite{40} & 280 &96.28\%&SGO-NN\\
CCANet \cite{42} & 280 &97.92\%&SGO-NN\\
\bf{DDCCANet} \cite{43} & 280 & \bf{98.50\%}&SGO-NN\\
\hline
\hline
\end{tabular}
\end{table}
\subsection{\emph{Object Recognition-The ETH-80 Database.}}
As a popular data set for multi-view feature representation studies, the ETH-80 database consists of 3280 color RGB object samples. All samples are divided into eight classes, including apples, cars, cows, cups, dogs, horses, pears and tomatoes. In our experiments, all samples are normalized at a size of 64 $\times$ 64 pixels. Moreover, 1640 images are randomly chosen to construct the training subset while the remaining images are used in testing. The raw data (R and G sub-channel images) is adopted as the two inputs for SGO-NN models.
%as shown in Figure. 6.
For comparison, recognition rates by SOTA methods in all three categories are listed in \textbf{Table 3}.\\
%\begin{figure}[H]
%\centering
%\includegraphics[height=1.0in,width=3.0in]{rg.eps}\\ Figure. 6 The two different views of raw data (R and G sub-channel images) on the ETH--80 database.
%\end{figure}
\vspace*{-10pt}
\begin{table}[h]
\small
\renewcommand{\arraystretch}{1.0}
\caption{\normalsize{Recognition accuracy on the ETH--80 database}}
\setlength{\abovecaptionskip}{0pt}
\setlength{\belowcaptionskip}{10pt}
\centering
\tabcolsep 0.05in
\begin{tabular}{cccc}
\hline
\hline
Methods & Training \#  & Accuracy & Type\\
%\hline
%\hline
%LapMCC \cite{37} & 984 & 69.50\%&W-I-ML\\
%HesMCC \cite{38} & 984 &73.52\%&W-I-ML\\
%CCANet \cite{18} & 500 &87.20\%&SML-NN\\
%\bf{DDCCANet} \cite{19} & 500 & \bf{88.18\%}&SML-NN\\
%\hline
%\hline
%ALP-TMR \cite{39} & 1400 &88.92\% &FFNN-DL\\
%CCANet \cite{18} & 1000 &91.97\%&SML-NN\\
%\bf{DDCCANet} \cite{19} & 1000 & \bf{92.24\%}&SML-NN\\
\hline
\hline
%SDNN \cite{41} & 1640 &82.80\%\\
%MMFML-M3 \cite{42} & 1640 & 92.53\%\\
%MKDR-LR \cite{43} & 1640 &91.20\%\\
%Kernel-LP \cite{44} & 1600 &89.11\% \\
%JDRML \cite{46} & 1640 &94.00\%\\
%MCCM-LE \cite{47} & 1640 &93.30\%\\
%Sparsity+Intra-task \cite{49} & 1640 & 92.50\%\\
SSL-TR \cite{56} & 1640 &93.40\%&WO-I-ML\\
%PLSRGStO \cite{67}  & 1640 &92.50\%&WO-I-ML\\
SRC+DPC \cite{57} & 1640 & 94.00\%&WO-I-ML\\
SML \cite{58} & 1640 &94.02\%&WO-I-ML\\
RMML \cite{78} & 1640 &94.25\%&WO-I-ML\\
%RMML-GM \cite{54} & 1640 & 90.25\%\\
%LRRTDR \cite{55} & 1640 & 92.00\%\\
%rBDLR \cite{56} & 2624 & 93.55\%&W-I-ML\\
%PDOD+AOD+KNN \cite{57} & 3239 & 93.25\%\\
TLRDA+PCA \cite{59} & 1640 & 92.00\%&C-NN\\
CMCM \cite{60} & 1640  &92.50\% &C-NN\\
Fine-tuned AlexNet \cite{61} & 1640 & 94.20\%&C-NN\\
CCANet \cite{42} & 1640 &93.98\%&SGO-NN\\
\bf{DDCCANet}\cite{43} & 1640 & \bf{94.40\%}&SGO-NN\\
\hline
\hline
\end{tabular}
\end{table}
\subsection{\emph{Object Recognition-The Caltech 256 Database.}}
Caltech 256 database contains a varying set of illumination, movements, backgrounds, etc. The classes are hand-picked to represent a wide variety of natural and artificial objects in various settings. In the experiments, for fair comparison, the same settings used in other studies are adopted. Specifically, 60 images are chosen from each class as training samples. A relatively simple DNN architecture, VGG-19, is employed to extract DL-based features, which serve as the input to the SGO-NN models. The performance of SOTA methods is reported in \textbf{Table 4}.\\%Two fully connected layers fc6 and fc7 are used for extracting the two view-based features. Since the dimensional number of fc6 and fc7 is 4096, both of them are reshaped as two-dimensional maps in the size of 64 $\times$ 64 as the input of CCANet/DDCCANet.
\vspace*{-10pt}
\begin{table}[h]
\small
\renewcommand{\arraystretch}{1.0}
\caption{\normalsize{Recognition accuracy with other methods on the Caltech 256}}
\setlength{\abovecaptionskip}{0pt}
\setlength{\belowcaptionskip}{10pt}
\centering
\tabcolsep 0.05in
\begin{tabular}{cccc}
\hline
\hline
Methods & Training \# & Accuracy & Type\\
\hline
\hline
%CMFA-SR \cite{58} &7710 &71.44\% \\
%DGFLP \cite{59} &7710 &69.17\% \\
%BMDDL \cite{61} &7710 &59.30\% \\

%ISC-LG \cite{63} &7710 &50.62\% \\
%BLF-FV \cite{64} &7710 &51.42\% \\
%OCB-FV \cite{65} &7710 &53.15\% \\
%LLC-SVM \cite{66} &7710 &34.50\% \\

%ResNet152 \cite{68} &7710 &78.00\% \\
%LLKc \cite{73} &7710 &72.09\% &WO-I-ML\\
%SROSR \cite{74} &7710 &75.60\% &WO-I-ML\\
%LMCCA \cite{46} &7710 &76.52\% &WO-I-ML\\
%DMCCA \cite{47}  &7710 &80.32\% &WO-I-ML\\
%MVLS \cite{70} &7710 &84.23\% &W-I-ML\\
%NR \cite{71} &7710 &84.40\% &W-I-ML\\
%SWSS-VGG \cite{75} &7710 &73.56\% &FFNN-DL\\
%Hybrid1365-VGG \cite{76} &7710 &76.04\% &FFNN-DL\\
%ResFeats152 \cite{77}  &7710 &79.50\% &FFNN-DL\\
%TransTailor \cite{82}  &7710 &85.30\% &FFNN-DL\\
%HMML \cite{75}  &12850 &64.06\% \\
%DCS \cite{76} &13770 &64.18\% \\
%CCANet \cite{42} & 7710 &85.07\%&SGO-NN\\
%\textbf{DDCCANet} \cite{43} &7710 & \textbf{86.33\%} &SGO-NN\\
%\hline
%\hline
%DCS \cite{76} &15300 &69.86\% \\

%ISC-LG \cite{63} &15420 &55.76\% \\
%OCB-FV \cite{65} &15420 &59.03\% \\
%LLC-SVM \cite{66} &15420 &40.10\% \\
%SWSS-VGG \cite{67} &15420 &76.25\% \\
%ResNet152 \cite{68} &15420 &81.90\% \\
%ResFeats152 \cite{69}  &15420 &82.10\% \\
%NAC \cite{78} &15420 &84.10\% \\
CMFA-SR \cite{62} &15420 &76.31\% &WO-I-ML\\
LLKc \cite{63} &15420 &75.36\% &WO-I-ML\\
%SMNN+VGG-19 \cite{64} &15420 &81.90\% &FFNN-DL\\
%SMNN+Resnet152 \cite{64} &15420 &82.30\% &FFNN-DL\\
Joint fine-tuning \cite{65} &15420 &83.80\% &C-NN\\
SMNN+Xception \cite{64} &15420 &84.70\% &C-NN\\
TransTailor \cite{66}  &15420 &87.30\% &C-NN\\
CCANet \cite{42} & 15420 &87.82\%&SGO-NN\\
\textbf{DDCCANet} \cite{43} &15420 & \textbf{88.34\%} &SGO-NN\\
\hline
\hline
\end{tabular}
\end{table}\\\indent
The three visual examples clearly show that the mathematically plausible DDCCANet brings out the discriminant information from the given data sets, resulting in superior performance compared with SOTA algorithms as shown in \textbf{Tables 2--4}.\\\indent In summary, the experimental results illustrated in \textbf{Tables 1--4} clearly show that both I-ML branches operate well in the experiments, with SGO-NN having a slight edge in the three visual examples and being substantially better in cross-model recognition. The results evidently justify the necessity of incorporating interpretability in ML research. %In addition, by integrating the SGO principles and NN architecture, not only more abstract and robust semantics are explored by NN structure, but also mathematical vigor and meaningful interpretation of the functionality of the SGO solutions are put into perspective, properly justifying its superior performance over the other categories of algorithms (e.g., WO-I-ML and FFNN-DL).
\section{V. DISCUSSION AND FUTURE PROSPECTS}
\subsection{Discussions}
In this subsection, we first discuss and emphasize on the key points raised in this paper:
\begin{enumerate}
  %\item NN-based ML methods have become a mainstream approach in processing numerous multimedia content such as text, images, audio, videos, and graphs. However, there still lack reasonable explanations about the underlying mechanism and behaviors due to the black-box nature.

  %\item  To address the black-box problem, studies on NN interpretability have attracted enormous attention from researchers and practitioners in the ML community and closely related fields, including multimedia. The ultimate purpose of studying model interpretability is to modify a NN model so that each convolutional layer generates interpretable knowledge. We have witnessed a great number of solutions proposed and investigated, with most being based on FFNN or pure DL-based strategy.
  \item I-ML gives ML models the ability to explain or to present their behaviors in understandable terms to humans, which better serves human beings and brings benefits to our society. As a result, there is a growing interest in both the academic and industrial sectors in I-ML and insight is being gained into working mechanisms of this class of ML models, both classical NN based I-ML and inherently I-ML.
  \item The investigation of the inherently I-ML models opens up a new front to address various challenges in model interpretability of ML with the objective of minimizing the black-box problem from the network design stage. This class of models obeys structural knowledge of the domain and can, at least, be partially justified by theoretical analysis such as physics laws and/or mathematical formulas. The current success and ongoing trend show that these models are expected to have a bright future in ML research and applications.
  \item Based on the K-A theorem/K-V UA theory, neurobiological signal processing facts and SGO principles, the SGO-NN models have demonstrated great promise in addressing the interpretability problem associated with contemporary ML, especially that pertinent to multi-modal image and multimedia analysis and recognition. In these models, not only are more abstract and robust semantics being explored by NN structure, but also mathematically meaningful interpretations of the functionality of the SGO solutions are put into perspective, vigorously justifying its superior performance over the other categories of algorithms (WO-I-ML and C-NN).
  \item Although SGO-NN is closely aligned with information fusion, the ETH-80 example with image pixels as network input demonstrated its power to handle the complete processing pipeline with raw data as input. Thus it can serve both as an information integrator of high level features and an end-to-end processor like the DNNs.
  \item Profiting from the parallel computing power of GPU, the most recent study \cite{68} shows that the calculation time of SGO-NN models has been greatly shortened, making it practically plausible in both academic research and real-world tasks.

      %\vspace*{-10pt}
%\begin{table}[h]
%\small
%\renewcommand{\arraystretch}{1.2}
%\caption{\normalsize{Running time on the different database}}
%\setlength{\abovecaptionskip}{0pt}
%\setlength{\belowcaptionskip}{10pt}
%\centering
%\tabcolsep 0.07in
%\begin{tabular}{ccc}
%\hline
%\hline
%Database & Running Time (Original) & Running Time (GPU)\\
%\hline
%\hline
%ORL & 500 seconds &20 seconds\\
%ETH-80  & 5400 seconds &180 seconds\\
%Caltech256 & 2.6 $\times$ $10^5$ seconds &1.8 $\times$ $10^4$ seconds\\
%\hline
%\hline
%\end{tabular}
%\end{table}

      %According to the collected exemplar applications, the integration of SML principles with NN architecture has demonstrated its capability of transforming the NN-based models explanations into a new domain in which information is more clearly represented for various multimedia processing tasks.
\end{enumerate}

\subsection{Future Prospects}
This paper prompts us to contemplate further on future prospects of I-ML, especially in ML and the intelligent multimedia communities with the following proposed considerations:
\begin{enumerate}
    \item \textbf{Challenges in the Study of I-ML.}
      There are still numerous critical challenges to understanding the interpretability of ML, such as:\\
      (a) how to discover more effective solutions that can learn interpretability from heterogeneous ML-based models. Although the research on model interpretability is a hot topic, it is still far from being thoroughly studied, especially when dealing with limited training data. It is our humble opinion that extracting complementary features via heterogeneous models is one of the potential solutions to this issue. It is known that information fusion is capable of exploiting complementary and/or consistency properties amongst individual features for effective knowledge discovery in multimedia computing \& synthesis, presenting a solution to the limited training data problem. Since heterogeneous models usually possess distinct architectures/algorithms, more effort should be devoted to study this class of ML models.\\
      (b) how to reveal the model interpretability when the input data contains temporal information (such as in video-based applications), an area which has not been extensively explored using I-ML models. For example, temporal features such as skeletons are extracted and combined with static features of color and depth to improve performance in action recognition. Hence, the demand of simultaneously and collaboratively processing static and dynamic information streams presents new challenges to the exploration and design of more powerful I-ML models.\\ \hspace*{\fill}

   %\item \textbf{Model interpretability based on information fusion strategy.}

   %The research work on DNNs interpretability has become more popular, though still far from being thoroughly studied, especially when dealing with limited training data (e.g., medical applications). It is our humble opinion that complementarity or heterogeneous feature extraction via distinct multimodal neural networks, the intrinsic relationship of different data generated by DNN, and the coupling law between multimodal DNNs are some of the most worthwhile future research directions in model interpretability. \\ \hspace*{\fill} \\

   \item \textbf{Model interpretability based on methodology fusion.}
   From both the survey and the exemplar applications presented in this paper, fusion methods integrating SGO principles and NN architecture are a promising branch in the study of I-ML, opening a door for the development of inherently interpretable learning models. However, most of the current SGO-NN based methods are only able to handle one or two data streams. To work with data from three or more sources, a situation frequently encountered in the real world, there is a natural demand to design new models/algorithms. We anticipate that the study of methodology fusion will continue maximizing the pros and minimizing the cons of SGO-NN models. %Another point particularly worth noting is that the combination of NN's ability for feature extraction in big data and SML's ability for mathematically vigorous coding and transformation with clear interpretation (stand alone or incorporating with NN), forms an unique and effective platform for NN interpretability.
\end{enumerate}

\section{VI. Conclusions}
This paper reviews recent advances and contemplates the future prospects of ML interpretability. %According to the current research, solutions are categorized into two classes: 1) Exploring interpretable ML by FFNN or pure DL-based methods; and 2) Exploring interpretable ML by integration of SGO principles with NN architecture.
%From the discussions at an analytical evaluation and comparison of model interpretability at the application level, this paper evidently demonstrates that: 1) the FFNN or pure DL-based methods provide an effective solution to the black-box nature in NN by revealing the model interpretability. 2) compared to the first category, the SGO-NN branch has shown at least a slight edge in terms of quality and performance. 3) the collaboration of SGO principles and NN architecture, especially with the availability of multi-view or multi-modal data, forms a unique and effective platform for interpretability of NN-based models, where SGO principles offer vigorous mathematical explanations.\\\indent
Performance and comparison on the collected exemplar applications using basic NNs/DNNs and the two categories of I-ML methods indicate that I-ML methods had evidently led to performance gains. In addition, methodology fusion of SGO principles and NN architecture brings about the benefit of reduced computing cost. From the survey paper, we reckon that more efforts should be put into the investigation of the study of I-ML.

\begin{IEEEbiography}{Lei Gao}{\,}is currently with the Department of Electrical, Computer, and Biomedical Engineering, Toronto Metropolitan University. His research interests include multimedia computing, multimodal information fusion, machine learning, and patter recognition. Lei Gao received his Ph.D. degree in the Ryerson University. He is a member of IEEE and a member of ACM. Contact him at iegaolei@gmail.com.
\end{IEEEbiography}

\begin{IEEEbiography}{Ling Guan}{\,}is currently a Professor with the Department of Electrical, Computer, and Biomedical Engineering, Toronto Metropolitan University. He has published extensively in multimedia processing and communications, human centered computing, machine learning, image and signal processing, and multimedia computing in the immersive environment. Ling Guan received the Ph.D. degree from the University of British Columbia. He is a Fellow of IEEE, an Elected Member of the Canadian Academy of Engineering. Contact him at lguan@ee.torontomu.ca.
\end{IEEEbiography}

\end{document}